\documentclass[twocolumn,aps,prl,superscriptaddress]{revtex4}

\usepackage{amsmath}
\usepackage{graphicx}
\usepackage{MnSymbol}

\begin{document}

\title{Physics of suction cups in air and in water}

\author{A. Tiwari}
\affiliation{PGI-1, FZ J\"ulich, Germany, EU}
\author{B.N.J. Persson}
\affiliation{PGI-1, FZ J\"ulich, Germany, EU}
\affiliation{www.MultiscaleConsulting.com}

\begin{abstract}
We present experimental results for the dependency of the pull-off time (failure time) 
on the pull-off force for suction cups in the air and in water. The results are
analyzed using a theory we have developed for the contact between suction cups and
randomly rough surfaces. The theory predicts the dependency of
the pull-off time (failure time) on the pull-off force, and is tested with measurements
performed on suction cups made from a soft polyvinyl chloride (PVC). 
As substrates we used sandblasted poly(methyl methacrylate) (PMMA).
The theory is in good agreement with the experiments in air, except for
surfaces with the root-mean-square (rms) roughness below $\approx 1 \ {\rm \mu m}$, 
where we observed lifetimes much longer than predicted by the theory.
We show that this is due to out-diffusion of plasticizer from the soft PVC, which block the 
critical constrictions along the air flow channels. In water some deviation between theory and
experiments is observed which may be due to capillary forces. We discuss the role of cavitation for the
failure time of suction cups in water.
\end{abstract}

\maketitle

\setcounter{page}{1}
\pagenumbering{arabic}




{\bf 1 Introduction}

All solids have surface roughness which has a huge influence on a large number of physical
phenomena such as adhesion, friction, contact mechanics and the leakage of seals\cite{Ref1,Ref2,Ref3,Ref4,Ref5,Ref6,Ref7,Ref8,Avi,BP,Creton,Gorb4,Mark}.
Thus when two solids with nominally flat surfaces are squeezed into contact, unless the applied
squeezing pressure is high enough, or the elastic modulus of at least one of the solids low enough,
a non-contact region will occur at the interface. If the non-contact region percolate
there will be open channels extending from one side of the nominal contact region to the other side.
This will allow fluid to flow at the interface from a high fluid pressure 
region to a low pressure region. 

For elastic solids with randomly rough surfaces the contact area percolate when the 
relative contact area $A/A_0 \approx 0.42$ (see \cite{Dapp}), where $A_0$ is the nominal 
contact area and $A$ the area of real contact (projected on the $xy$-plane). When the
contact area percolate there is no open (non-contact) channel at the interface 
extending across the nominal contact region,
and no fluid can flow between the two sides of the nominal contact.

The discussion above is fundamental for the leakage of static seals\cite{seal1,seal2,seal3,seal4}. 
Here we are interested in rubber suction cups. In this application, the contact between the suction cup 
and the counter surface (which form an annulus) must be so tight that negligible
fluid can flow from outside the suction cup to inside it. 

Suction cups find ubiquitous usage in our everyday activities such as hanging of items to smooth surfaces in our 
houses and cars, and for technologically demanding applications such as lifting fragile and heavy objects safely in a controlled manner using suction cups employing vacuum pumps. Suction cups are increasingly used in robotic applications, such as 
robots which can climb walls and clean windows. The biomimetic design of suction cups 
based on octopus vulgaris, remora (sucker fish), limpets and Northern Clingfish is an
 area of current scientific investigations whose main objectives is to manufacture suction 
	cups exhibiting adhesion under water and on surfaces with varying degree of surface 
	roughness\cite{Dit,Sand}. Recently Iwasaki et. al. have presented the concept of 
	magnet embedded suction cups for in-vivo medical applications\cite{Iwas}.

\begin{figure}
\includegraphics[width=0.35\textwidth,angle=0]{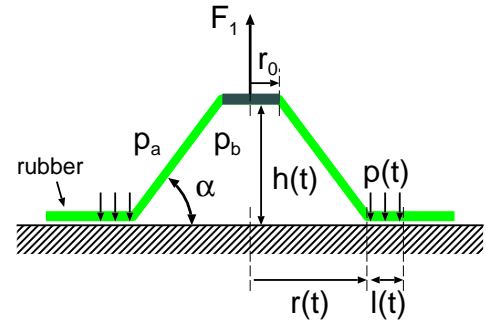}
\caption{\label{pic.eps}
Schematic picture of the suction cup pull-off experiment used in the present study.
The container with the suction cup is either empty or filled with distilled water.
}
\end{figure}

\vskip 0.1cm
{\bf 2 Theory: gas leakage}

The suction cups we study below can be approximated as a truncated cone with the diameter $2r_1$. The angle $\alpha$ 
and the upper plate radius $r_0$ are defined in Fig. \ref{pic.eps}.
When a suction cup is pressed in contact with a flat surface the rubber cone
will make apparent contact with the substrate in an annular region, but the contact pressure will be largest 
in a smaller annular region of width $l(t)$ formed close to
the inner edge of the nominal contact area (see Fig. \ref{pic.eps}). 
We will assume that the rubber-substrate contact pressure in this region of space is constant, $p=p(t)$, and zero elsewhere. 

If we define $h_0 = r_0 {\rm tan} \alpha$ the volume of gas inside the suction cup is
$$V= \pi r^2 {1\over 3} (h+h_0) - \pi r_0^2 {1\over 3} h_0$$
Since 
$${r\over r_0} = 1+{h \over h_0}\eqno(1)$$
we get
$$V = V_0 \left [\left ({r\over r_0}\right )^3-1\right ]\eqno(2)$$
where $V_0 = \pi r_0^2 h_0 /3$. 
The elastic deformation of the rubber film (cone) needed to make contact 
with the substrate require a normal force $F_0 (h)$, which we will refer
to as the cup (non-linear) spring-force. This force result from the bending of the film 
and to the (in-plane) stretching and compression of the film
needed to deform (part of) the conical surface into a flat circular disc or annulus.
The function $F_0(h)$ can be easily measured experimentally (see below). 

We assume that the rubber cup is in repulsive contact with the substrate
over a region of width $l(t)$. Since the thickness of the suction cup material decreases as $r$ increases, we expect that $l$
decreases as $r$ increases. From optical pictures of the contact we have found that to a good approximation
$$l \approx l_0 +l_{\rm a} \left (1-{r\over r_0}\right ) = l_0 - l_a {h\over h_0}\eqno(3)$$
where $l_a = (l_1-l_0)/(1-r_1/r_0)$ where $l_1$ is the width of the contact region when $r=r_1$, 
and $l_0$ the width of the contact region when $r=r_0$.
The contact pressure $p=p(t)$ in the circular contact strip is assumed to be constant
$$p \approx { F_0(h) \over 2 \pi r l}+\beta (p_{\rm a}-p_{\rm b})\eqno(4)$$
where $\beta$ is a number between 0 and 1, which takes into account that the gas pressure 
(in the non-contact region) in the strip
$l(t)$ changes from $p_{\rm a}$ for $r>r(t)+l(t)$ to $p_{\rm b}$ for $r<r(t)$, while the outside pressure 
is $p_{\rm a}$. 

Assume that the pull-force $F_1$ act on the suction cup (see Fig. \ref{pic.eps}). The sum of $F_1$ and the
cup spring-force $F_0$ must equal the force resulting from the pressure difference
between outside and inside the suction cup,  i.e.
$$F_0+F_1 = \pi r^2  \left (p_{\rm a} - p_{\rm b}\right )\eqno(5)$$
We assume that the air can be treated as an ideal gas so that
$$p_{\rm b} V_{\rm b} = N_{\rm b} k_{\rm B}T.\eqno(6)$$

The number of molecules per unit time entering the suction cup, $\dot N_{\rm b} (t)$, is given by
$$\dot N_{\rm b} = f(p,p_{\rm a},p_{\rm b}) {L_y\over L_x}\eqno(7)$$
Here $L_x$ and $L_y$ are the lengths of the sealing region along and orthogonal to the gas leakage direction, respectively.
In the present case
$${L_y\over L_x} = {2 \pi r \over l}$$
The (square-unit) leak-rate function $f(p,p_{\rm a},p_{\rm b})$ will be discussed below.

The equations (1)-(7) constitute 7 equations from which the following 7 quantities can be obtained:
$h(t)$, $r(t)$, $l(t)$, $V(t)$, $p_{\rm b}(t)$, $p(t)$ and $N_{\rm b}(t)$.
The equations (1)-(7) can be easily solved by numerical integration. 

The suction cup stiffness force $F_0(h)$ depends on the speed with which the suction cup is compressed (or decompressed).
The reason for this is the viscoelastic nature of the suction cup material. To take this effect into account 
we define the contact time state variable $\phi (t)$ as\cite{Ruina,Rice,Ref1}:
$$\dot \phi = 1-\dot r \phi /l\eqno(8)$$
with $\phi(0) = 0$. 
For stationary contact, $\dot r = 0$, this equation gives just the time $t$ of stationary contact, $\phi (t) =t$.
When the ratio $\dot r/l$ is non-zero but constant (8) gives  
$$\phi (t) =  \left (1-e^{-t/\tau}\right ) \tau, $$
where $\tau = l/\dot r$. Thus for $t >> \tau$ we get $\phi (t) = \phi_0 = \tau$, which is the time a particular point on the suction cup surface
stay in the rubber-substrate contact region of width $l(t)$. It is only in this part of the rubber-substrate nominal contact region where a strong (repulsive)
interaction occur between the rubber film and the substrate, and it is region of space which is most important for the gas sealing process.

From dimensional arguments we expect that $F_0(h)$ is proportional to the effective elastic modulus of the cup material.
We have measured $F_0(h)$ at a constant indentation speed $\dot h$, corresponding to a constant radial velocity $\dot r = \dot h (r_0/h_0)$
(see (1)). In this case the effective elastic modulus is determined by the relaxation modulus
$E_{\rm eff} (t)$ calculated for the contact time $\phi_0 = l/\dot r$. 
However, in general $\dot r$ may be strongly time-dependent. We can take that into account by 
replacing the measured $F_0(h)$ by the function $F_0(h) E_{\rm eff} (\phi(t))/E_{\rm eff} (\phi_0)$.

\begin{figure}
        \includegraphics[width=0.4\textwidth]{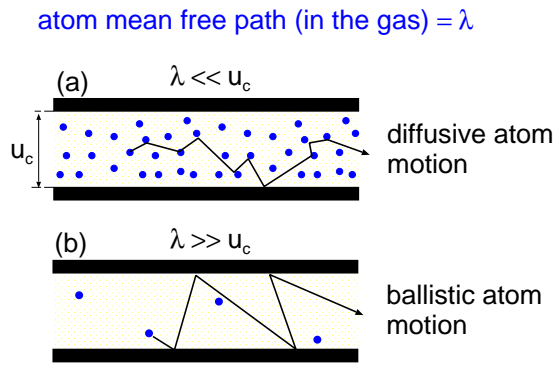}
        \caption{\label{Ballistic.eps}
Diffusive (a) and ballistic (b) motion of the gas atoms in the critical junction.
In case (a) the gas mean free path $\lambda$ is much smaller than the gap width $u_{\rm c}$ and the gas
molecules makes many collisions with other gas molecules before a collision with the solid walls.
In the opposite limit, when $\lambda >> u_{\rm c}$ the gas molecules makes many collisions with the solid wall before colliding
with another gas molecule. In the first case (a) the gas can be treated as a
(compressible) fluid, but this is not the case in (b).}
\end{figure}

\vskip 0.1cm
{\bf 2.1 Diffusive and ballistic gas leakage}

The gas leakage result from the open (non-contact) channels at the interface between the rubber
film and the substrate. Most of the leakage occur in the biggest open flow channels. 
The most narrow constriction in the biggest open channels are denoted as the
critical constrictions. Most of the gas pressure drop occur over the critical constrictions,
which therefore determine the leak-rate to a good approximation.
The surface separation in the critical constrictions is denoted by $u_{\rm c}$. Theory shows that the lateral
size of the critical constrictions is much larger than the surface separation $u_{\rm c}$ (typically by
a factor of $\sim 100$)\cite{seal1,seal2,seal3}. 

In the theory for suction cups enters the leak-rate function $f(p,p_{\rm a},p_{\rm b})$ (see (7)).
This function can be easily calculated when the gas flow through the critical constrictions occur
in the diffusive and ballistic limits (see Fig. \ref{Ballistic.eps}). Here we present 
an interpolation formula which is (approximately) valid
independent of the ratio between the gas mean free path and the surface separation at the critical constrictions:
$$\dot N_{\rm b} = {1  \over 24} {L_y\over L_x} {(p_{\rm a}^2 -p_{\rm b}^2 )\over k_{\rm B}T} 
{u_{\rm c}^3  \over \eta }\left (1+12 {\eta \bar v \over (p_{\rm a}+p_{\rm b}) u_{\rm c}} \right )\eqno(9)$$
Here $\eta$ is the gas viscosity and $k_{\rm B}T$ the thermal energy.
The gas leakage equation (9) is in good agreement with treatment using the Boltzmann equation, and with experiments\cite{Boltz1,tobe}.
To calculate $u_{\rm c}$ we need the relation between the interfacial separation $u$ and the
contact pressure $p$. For this we have used the Persson contact mechanics theory\cite{BP,Alm,uc1}.

\vskip 0.1cm
{\bf 3 Experimental }

We carried out the leakage experiments in air and water using two suction cups, denoted A and B, made from soft PVC.
These suction cups have different geometrical designs, which has a  influence on the suction cup stiffness
and failure time, as discussed in section 3.2 and 4.1 respectively.

\begin{figure}
	\includegraphics[width=0.45\textwidth]{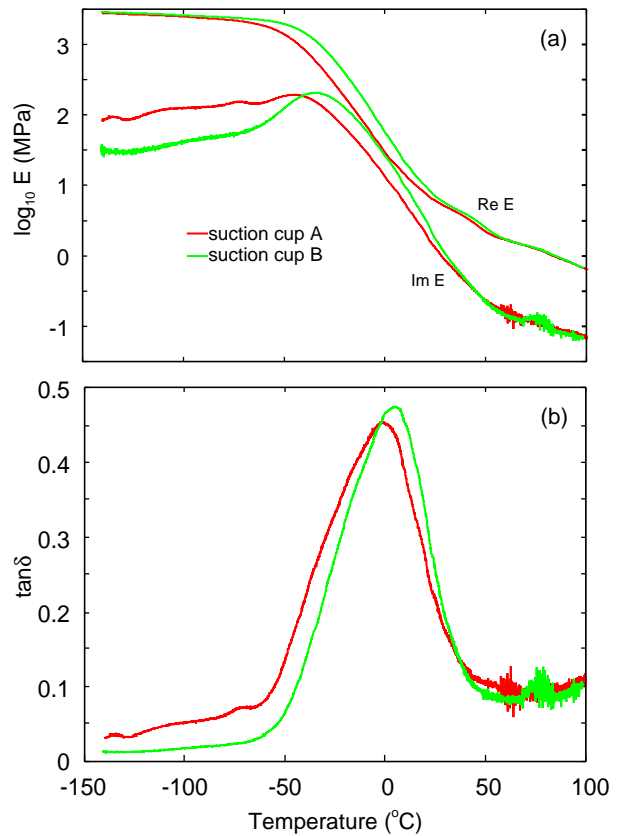}
	\caption{\label{ModlusVsTemperature.and.tand.eps}
		The dependency of the low strain ($\epsilon = 0.0004$) viscoelastic modulus (a), and ${\rm tan}\delta = {\rm Im}E/{\rm Re}E$ (b), on the temperature for the frequency $f=1 \ {\rm Hz}$. For the soft PVC of suction cup A (red) and B (green).}
\end{figure}

\vskip 0.1cm
{\bf 3.1 Viscoelastic modulus of PVC}

Viscoelastic modulus measurements of the suction cups A and B where carried out 
in oscillatory tension mode using a Q800 DMA instrument produced by TA Instruments.
Fig. \ref{ModlusVsTemperature.and.tand.eps} shows (a) the temperature dependency of the low strain 
($\epsilon = 0.0004$) modulus $E$, and (b) ${\rm tan}\delta = {\rm Im}E/{\rm Re}E$, 
for the frequency $f=1 \ {\rm Hz}$. Results are shown for the soft PVC of suction cup A (red lines) and B (green lines).
Note that both materials exhibit very similar viscoelastic modulus.
If we define the glass transition temperature as the temperature where
${\rm tan}\delta$ is maximal (for the frequency $f=1 \ {\rm Hz}$) then $T_{\rm g} \approx 0^\circ {\rm C}$.

\begin{figure}
	\includegraphics[width=0.45\textwidth]{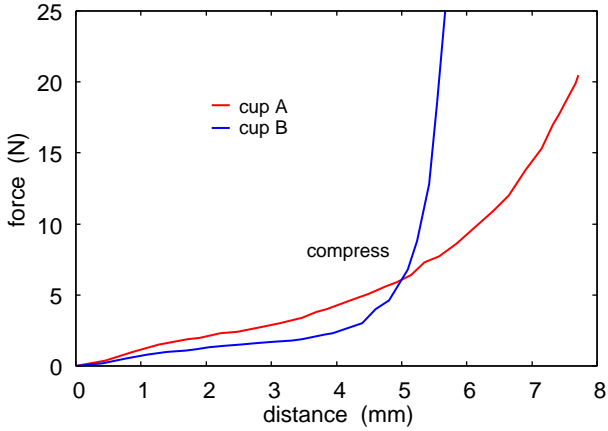}
	\caption{\label{1h.2F0.old.new.cup.eps}
		The stiffness force $F_0(h)$ (in N) as a function of the squeezing (or compression) 
distance (in mm) for the suction cups A (red) and B (blue).
		The suction cups are squeezed against a smooth glass plate with a hole 
in the center through which the air can leave so the air pressure
		inside the rubber suction cup is the same as outside (atmospheric pressure). The glass plate is lubricated with soap-water.}
\end{figure}

\vskip 0.1cm
{\bf 3.2 Suction cup stiffness force $F_0$}

We have measured the relation between the normal force $F_0$ and the normal displacement 
of the top of a suction cup.
In the experiments we increase the displacement of the top plate (see Fig. \ref{pic.eps}) 
at a constant speed and measure the resulting force. We show results for two different suction cups, denoted A and B.

We have measured the force $F_0$ for the suction cups squeezed against a smooth glass plate lubricated by
soap water. The glass plate has a hole below the top of the suction cup; this allowed the air to leave the suction cup
without any change in the pressure inside the suction cup (i.e. $p_{\rm b} = p_{\rm a}$ is equal to the atmospheric
pressure). Fig. \ref{1h.2F0.old.new.cup.eps} shows the stiffness force $F_0(h)$ (in N) as a function of the 
squeezing (or compression) distance (in mm) for the suction cups A (red) and B (blue).

The suction cups A and B are both made from similar type of soft PVC and both have the diameter $\approx 4 \ {\rm cm}$.
However, for suction cup B the angle $\alpha = 21^\circ$ in contrast to $\alpha = 33^\circ$ for suction cup A, and 
the PVC film is thicker for the cup A.
This difference in the angle $\alpha$ and the film thickness influence
the suction cup stiffness force as shown in Fig. \ref{1h.2F0.old.new.cup.eps}. Note that before the strong increase in the $F_0(h)$ curve
which result when the suction cup is squeezed into complete contact with the counter surface 
the suction cup A has a stiffness nearly twice as high as that of the suction cup B.

\begin{figure}
	\includegraphics[width=0.45\textwidth]{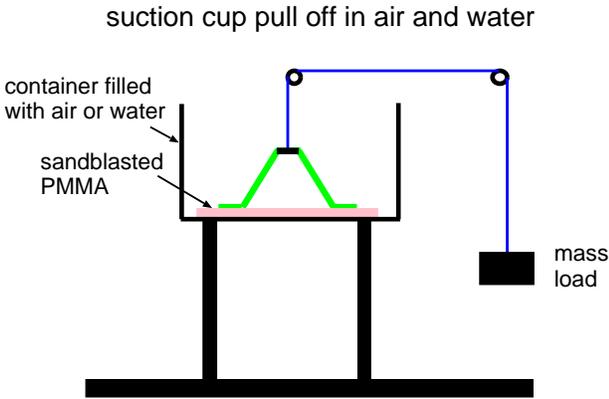}
	\caption{\label{picE.eps}
Schematic picture of the experimental set-up for measuring the pull-off force in the air and in water.
}
\end{figure}

\vskip 0.1cm
{\bf 4 Experiment: Failure of suction cup in air}

We have studied how the failure time of a suction cup depends on the pull-off force (vertical load) and
the substrate surface roughness (see Fig. \ref{picE.eps}). 
The suction cup was always attached to the lower side of a horizontal surface and a mass-load
was attached to the suction cup. We varied the mass-load from 0.25 kg to 8 kg.
If full vacuum would prevail inside the suction cup, the maximum possible pull-off force would be $\pi r_1^2 p_{\rm a}$.
Using $r_1 = 19 \ {\rm mm}$ and $p_{\rm a} = 100 \ {\rm kPa}$ we get $F_{\rm max} = 113 \ {\rm N}$ or about
11 kg mass load. However, the maximum load possible in our experiments for a smooth substrate surface 
is about 9 kg, indicating that no complete vacuum was obtained. This may, in least in part, 
be due to problems to fully remove the air inside the suction cup in the initial state. In addition we have
found that for mass loads above $8 \ {\rm kg}$ the pull-off is very sensitive 
to instabilities in the macroscopic deformations of the suction cup, probably
resulting from small deviations away from the vertical direction of the applied loading force. 

\begin{figure}
        \includegraphics[width=0.45\textwidth]{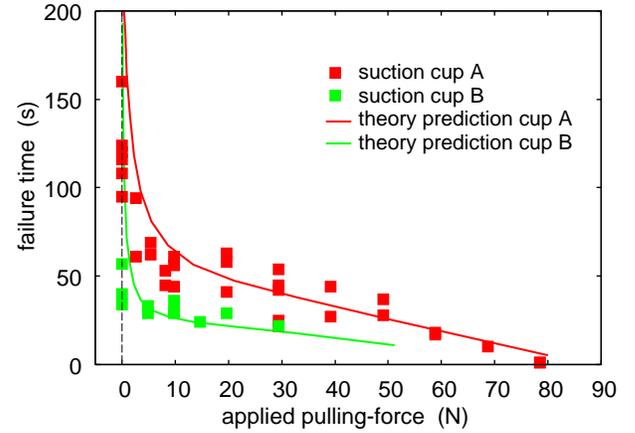}
        \caption{\label{1LoadingForce.2lifetime.oldRED.newGREEN.eps}
The dependency of the pull-off time (failure time) on the applied (pulling) force. The soft PVC suction cups A and B
are in contact with a sandblasted PMMA surface with the rms roughness $1.89 \ {\rm \mu m}$. All surfaces were cleaned with
soap water before the experiments.}
\end{figure}

\begin{figure}
        \includegraphics[width=0.45\textwidth]{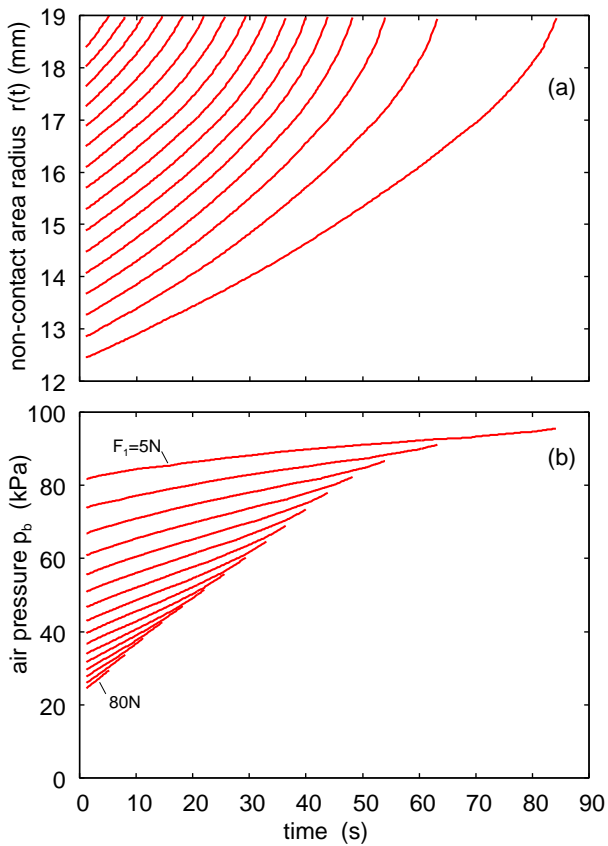}
        \caption{\label{1time.2r.3pa.x.eps}
The calculated dependency time dependency of the (a) radius of the non-contact region and (b) the pressure in the suction cup, 
for several pull-off forces (from $F_1=5 \ {\rm N}$ in steps of $5 \ {\rm N}$
to $80 \ {\rm N}$). The soft PVC suction cup is in contact with a sandblasted PMMA surface with the rms roughness $1.89 \ {\rm \mu m}$. 
}
\end{figure}

\vskip 0.1cm
{\bf 4.1 The dependency of the failure time on the pull-off force}

Fig. \ref{1LoadingForce.2lifetime.oldRED.newGREEN.eps} 
shows the dependency of the pull-off time 
(failure time) on the applied pulling force. The results are for the soft PVC suction cups A (red) and B (green)
in contact with a sandblasted PMMA surface with the rms-roughness $1.89 \ {\rm \mu m}$. 
Before the measurement, all surfaces were cleaned with
soap water. 
The solid lines are the theory predictions, using as input the surface roughness power spectrum of the 
PMMA surface, and the measured stiffness of the suction cup, the latter corrected for viscoelastic time-relaxation
as described above. Note the good agreement between the theory and the experiments in spite of the simple nature
of the theory.

Fig. \ref{1time.2r.3pa.x.eps}(a) shows the calculated time dependency of the radius of the non-contact region, 
and (b) the gas pressure in the suction cup. We show results for several pull-off forces from $F_1=5 \ {\rm N}$ in steps of $5 \ {\rm N}$
to $80 \ {\rm N}$.  

The smaller angle $\alpha$ for suction cup B than for cup A imply that if the same amount of gas would leak into the suction cups the
gas pressure $p_{\rm b}$ inside the suction cup will be highest for the suction cup B. This will tend to reduce the lifetime of cup B. Similarly,
the smaller stiffness of the cup B result in smaller contact pressure $p$,
which will increase the leakage rate and reduce the lifetime.
Hence both effects will make the lifetime of the suction cup B smaller than that of the cup A.

\begin{figure}
        \includegraphics[width=0.45\textwidth]{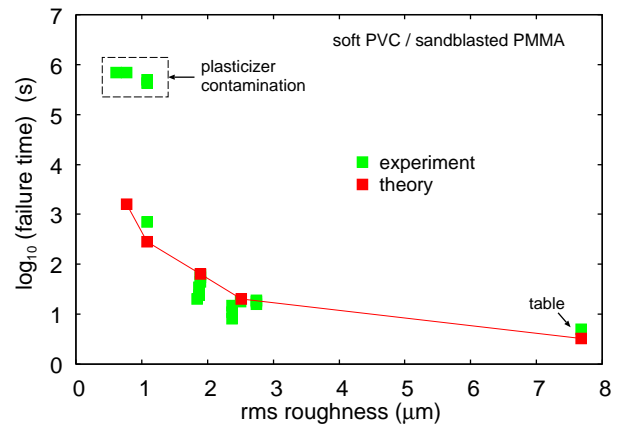}
        \caption{\label{1rms.2logFailureTime.eps}
The dependency of the pull-off time (failure time) on the substrate surface roughness. For soft PVC suction cups
in contact with sandblasted PMMA surfaces with different rms-roughness, and a table surface.
The pull-off force $F_1 = 10 \ {\rm N}$.
All surfaces were cleaned with soap water before the experiments, and a new suction cup was used for each experiment.}
\end{figure}

\vskip 0.1cm
{\bf 4.2 The dependency of the failure time on the surface roughness}

Fig. \ref{1rms.2logFailureTime.eps} shows the dependency of the pull-off time 
(failure time) on the substrate surface roughness. The results are for the type A soft PVC suction cups
in contact with sandblasted PMMA surfaces with different rms roughness, and a table surface.
Note that for ``large'' roughness the predicted failure time is in good agreement with the measured data, but for rms roughness
below $\approx 1 \ {\rm \mu m}$ the measured failure times are much larger than the theory prediction. 
In addition, the dependency of the radius $r(t)$ of the non-contact region on time is very different in the two cases: 
For roughness larger than $\approx 1 \ {\rm \mu m}$ the radius increases 
continuously with time as also expected from theory (see Fig. \ref{1time.2r.3pa.x.eps}(a)).
For roughness below $\approx 1 \ {\rm \mu m}$  the boundary line $r(t)$ stopped to move a short time after applying the pull-off force, 
and remained fixed until the detachment occurred by a rapid increase in $r(t)$ (catastrophic event).
We attribute this discrepancy between theory and experiments 
to transfer of plasticizer from the soft PVC to the PVC-PMMA interface; this (high viscosity) 
fluid will fill-up the critical constrictions
and hence stop, or strongly reduce, the flow of air into the suction cup. 
This is consistent with many studies\cite{plast2}
of the transfer of plasticizer from soft PVC to various contacting materials.
These studies show typical transfer rates
(at room temperature) corresponding to a $\sim 1-10 \ {\rm \mu m}$ thick film of plasticizer 
after one week waiting time. Optical pictures of the rough PMMA surface after long contact
with the suction cup A also showed darkened (and sticky) annular regions indicating transfer of material
from the PVC to the PMMA surface. 

\begin{figure}
	\includegraphics[width=0.45\textwidth]{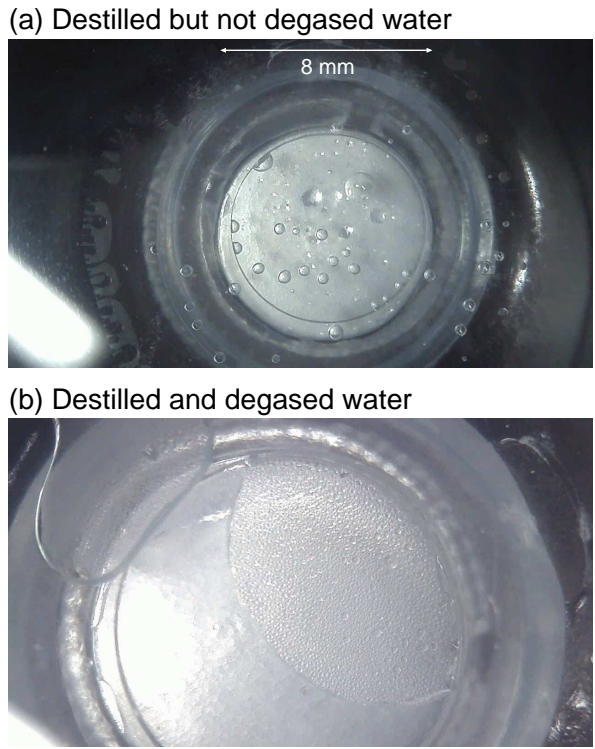}
	\caption{\label{air.eps}
Optical pictures of the center region of a suction cup after squeezing it against a smooth
glass plate in (a) distilled but not degassed water, and (b) distilled and degassed (at 0.2 atmosphere pressure) water.
In (a) we squeezed the cup with about $15 \ {\rm N}$ against the glass plate, and then pulled it with a similar force for $10 \ {\rm s}$ and 
then removed the pulling force and waited $\sim 5 \ {\rm minutes}$ before taking the picture shown. In (b) we did the same
but using a bigger squeezing and pulling force, about $60 \ {\rm N}$. In both cases
cavitation has occurred, and we observed a slow evolution in time of the gas covered region.
The central (or top) part of the suction cup is a flat circular metal disc (diameter $8 \ {\rm mm}$) covered by a PVC film.
The magnification in (b) is higher than in (a). Note the condensation of small water droplets on the glass surface in the
cavity region in (b). These droplets was growing with increasing time while the size of the gas bubble decreased slowly.
}
\end{figure}

\vskip 0.1cm
{\bf 5 Theory: liquid (water) leakage}

Assuming that the water can be treated as an incompressible Newtonian liquid, and assuming laminar flow, 
the volume of liquid flowing per unit time into the suction cup is given by 
$$\dot V_{\rm b} = {1  \over 12 \eta} {L_y\over L_x} u_{\rm c}^3   \left (p_{\rm a} -p_{\rm b} \right ). \eqno(10)$$ 
This equation replaces (9), which is valid for gas flow, but all the other equations are unchanged.

In the experiments reported on below, the suction cup
is squeezed vertically into contact with the counter surface (here a PMMA sheet). Even if no gas (air) can be detected
inside the suction cup before squeezing it in contact with the PMMA surface,
after removing the squeezing force we always observe a gas filled region at the top of the suction cup
(see Fig. \ref{air.eps}).
We believe this result from the spring force generating a reduced pressure inside the suction cup, which result in cavitation.
When loaded with a pull-off force this gas region expand and result in a pressure inside the suction cup which is somewhere between
atmospheric pressure and perfect vacuum (zero pressure). If all the air would have been removed, the pressure in the water
could, at least initially, be negative (below vacuum), where the liquid is under mechanical tension. 
In fact, pressures as low as $p_{\rm b} \approx - 20 \ {\rm MPa}$ has been observed
for water at short times\cite{Cav}. However, this state is only metastable and after long enough time one would
expect the nucleation of a gas bubble in the liquid, and the liquid pressure would increase above zero. In fact, water in thermal
(or kinetic) equilibrium with the normal atmosphere will have dissolved air molecules
(the volume ratio of dissolved gas (at atmospheric pressure) to water is about 0.04 at room temperature), 
and is unstable against cavitation whenever
the pressure falls below the atmospheric pressure. Thus, the problem of determining the pressure in the fluid occurring
inside the suction cup is nontrivial, and depend on trapped air bubbles and on how long time the reduced 
(compared to the atmospheric pressure) pressure prevail.  

When a suction cup is used in water,
when the pull-off force $F_1 \rightarrow 0$, at pull-off the suction cup contains water of atmospheric
pressure. When a suction cup is used in the normal atmosphere the suction cup is instead filled 
with air of atmospheric pressure. Assuming an
ideal gas the volume of air of atmospheric pressure flowing per unit time into the suction cup is determined by
$p_{\rm a} \dot V =\dot N_{\rm b} k_{\rm B}T$. Using (9) this gives
$$\dot V = {1  \over 24} {L_y\over L_x} {(p_{\rm a}^2 -p_{\rm b}^2 )\over p_{\rm a}} 
{u_{\rm c}^3  \over \eta }\left (1+12 {\eta \bar v \over (p_{\rm a}+p_{\rm b}) u_{\rm c}} \right )$$
$$={1  \over 12 \eta} {L_y\over L_x} u_{\rm c}^3   \left (p_{\rm a} -p_{\rm b} \right ) Q\eqno(11)$$
where
$$Q={p_{\rm a}+p_{\rm b}\over 2p_{\rm a}}
\left (1+12 {\eta \bar v \over (p_{\rm a}+p_{\rm b}) u_{\rm c}} \right )\eqno(12)$$
Thus the gas leakage rate differ from the liquid leak-rate by a factor $Q$, which is a product of a 
factor $(p_{\rm a}+p_{\rm b})/2p_{\rm a}$,
derived from the fact that the gas is a compressible fluid, and another factor arising from ballistic air flow. The latter
factor does not exist in the liquid case because of the short molecule mean free path in the liquid.

\begin{figure}
	\includegraphics[width=0.45\textwidth]{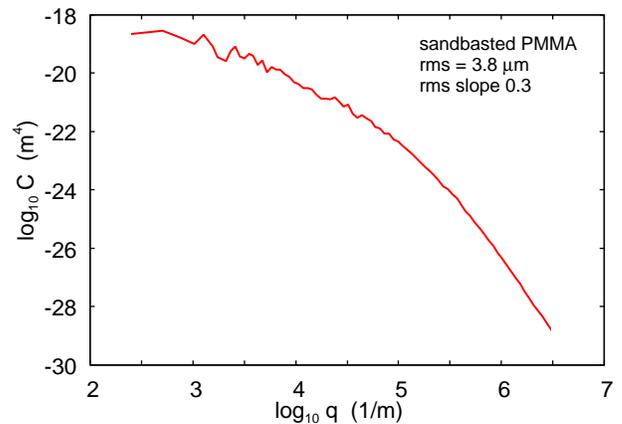}
	\caption{\label{PMMASandblasted10min8bar.eps}
		The surface roughness power spectrum of sandblasted PMMA. 
Sandblasting was carried out for 10 min at a pressure of 8 bars. The root mean square roughness is $3.8 \ {\rm \mu m}$.}
\end{figure}

\begin{figure}
	\includegraphics[width=0.45\textwidth]{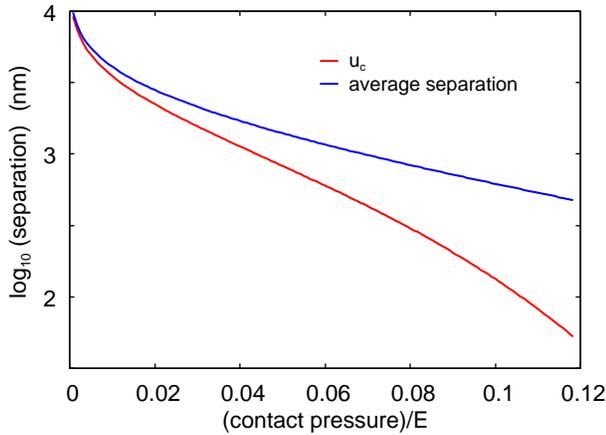}
	\caption{\label{1p.over.E.2uc.and.uaverage.eps}
	The average surface separation $\bar u$, and the separation $u_{\rm c}$ at the critical constriction, as a function of the 
nominal contact pressure in units of the modulus $E$. For the surface with the 
power spectrum given in Fig. \ref{PMMASandblasted10min8bar.eps}.}
\end{figure}

\begin{figure}
	\includegraphics[width=0.45\textwidth]{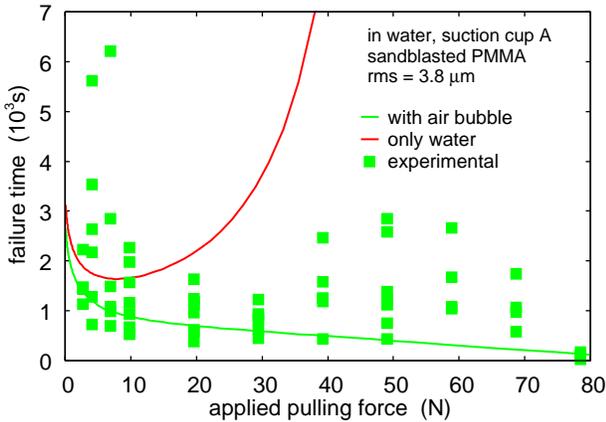}
	\caption{\label{fitsuctioncup.eps}
		Squares: The dependency of the failure time (time to pull-off) on the applied (pulling) force for the (soft 
PVC) suction cup  in contact with a sandblasted PMMA surface (with the rms roughness $3.8 \ {\rm \mu m}$) 
in water. The red and green solid lines are the theory predictions assuming that in the initial state, before applying the pull-off
force, there is some trapped water (but no air), or trapped air (but no water) 
inside the suction cup, so the initial radius of the detached region for no external load is about $10 \ {\rm mm}$. 
When only water occur, if the applied pull-force is big enough, a negative pressure (mechanical tension)
develop in the water which will pull the surfaces further together in the
contact strip $l(t)$. This reduces the water leakage and increases the failure time.
When air occur in the suction cup the pressure is always positive, but below the atmospheric pressure.}
\end{figure}

\begin{figure}
	\includegraphics[width=0.45\textwidth]{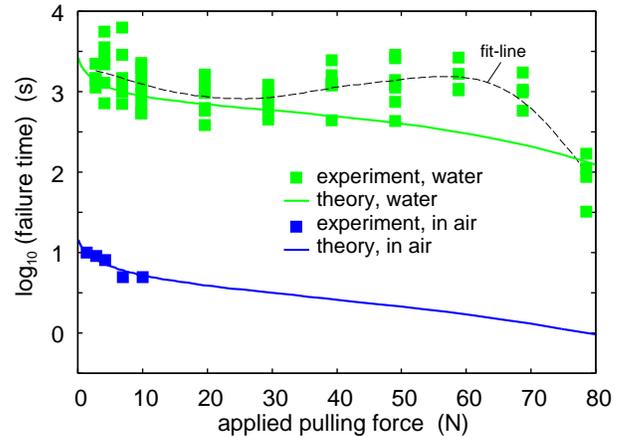}
	\caption{\label{1pulloffF.2LOGlifetime.eps}
		Squares: The dependency of the logarithm of the failure time (time to pull-off) on the applied (pulling) force for the 
(soft PVC) suction cup  in contact with a sandblasted PMMA surface (with the rms roughness $3.8 \ {\rm \mu m}$) 
in water (green) and in the air (blue). 
The green and blue solid lines are the theory predictions assuming that in the initial state, before applying the pull-off
force, there is trapped air inside the suction cup, giving an initial radius of the detached region for no external load 
of $\approx 10 \ {\rm mm}$ as observed experimentally. 
The black dashed line is a fit-line to the experimental data in water.
}
\end{figure}

\begin{figure}
	\includegraphics[width=0.45\textwidth]{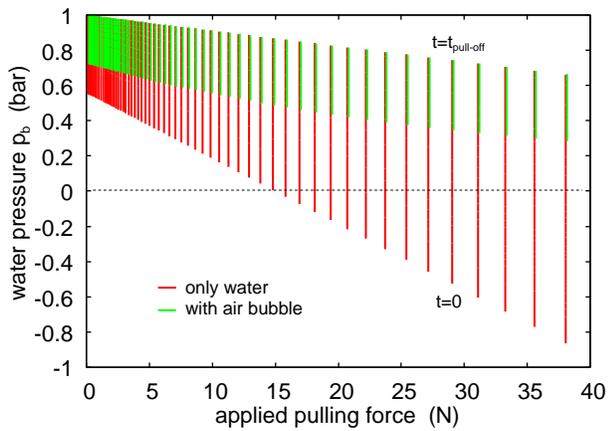}
	\caption{\label{1F1.2pb.1.eps}
The pressure $p_{\rm b}$ in the water inside the suction cup as a function of the applied
pulling force. The vertical lines give the water pressure as the time changes from start
of pull-off ($t=0$) to detachment ($t=t_{\rm pull-off}$). The red lines assumes only water inside the
suction cup while the green lines assumes a small air bubble trapped in the initial state. 
In the latter case the water (and air) pressure inside the suction cup is always above the vacuum pressure $p=0$,
while in the first case, when the pull-off force is bigger than
$\approx 15 \ {\rm N}$, the water pressure is negative for some initial time period.
}
\end{figure}

\begin{figure}
	\includegraphics[width=0.45\textwidth]{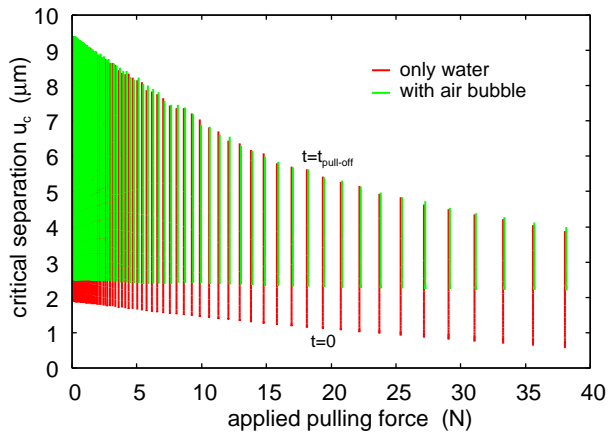}
	\caption{\label{1F1.2uc.1.eps}
The separation at the critical junction, $u_{\rm c}$, as a function of the applied
pulling force. The vertical lines give the critical separation as the time changes from start
of pull-off ($t=0$) to detachment ($t=t_{\rm pull-off}$). The red lines assumes only water inside the
suction cup, while the green lines assumes a small air bubble trapped in the initial state. 
The critical separation for short contact times is smaller in the former case owing to the lower
fluid pressure in the suction cup, which pull the surfaces in the nominal contact strip $l(t)$
closer together (see last term in (4)).
}
\end{figure}

\begin{figure}
	\includegraphics[width=0.45\textwidth]{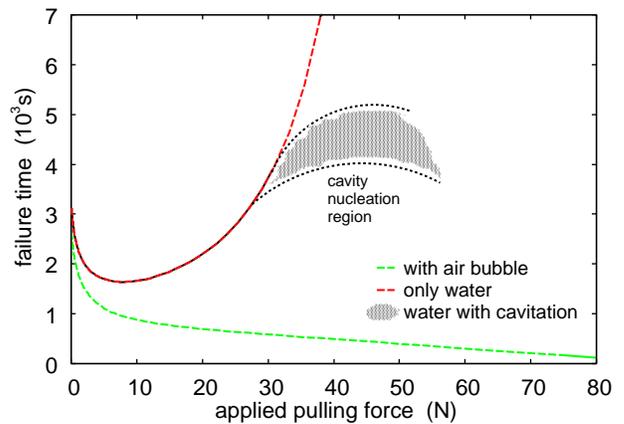}
	\caption{\label{1pullforce.2loglifetime.1.eps}
A schematic picture illustrating the influence of cavitation on the failure time.
For small applied pulling force the pressure in the water in the suction cup is not low enough to
induce cavitation, but for large enough pulling-force cavitation occur. In this case,
since cavitation is a stochastic process involving thermally activated nucleation of an air bubble, 
when the experiment is repeated many times large fluctuations in the failure time may occur (dotted region).
}
\end{figure}

\vskip 0.1cm
{\bf 6 Experiment: Failure of suction cup in water}

We have studied the failure time for suction cup A immersed in distilled water.
The experimental set up is shown in Fig. \ref{picE.eps}. Here, the suction cup is squeezed against the rough PMMA countersurface
under water. The water level is at least 20cm above the contacting interface. 
Fig. \ref{PMMASandblasted10min8bar.eps} shows the surface roughness power spectrum of the sandblasted PMMA 
surface used in water. The surface has the rms roughness amplitude $3.8 \ {\rm \mu m}$ and the rms slope
0.3. 

Using the power spectrum in Fig. \ref{PMMASandblasted10min8bar.eps} and the Persson contact 
mechanics theory, in Fig. \ref{1p.over.E.2uc.and.uaverage.eps}
we show the calculated average surface separation $\bar u$, and the separation $u_{\rm c}$ at the critical constriction, as a function of the 
nominal contact pressure in units of the modulus $E$ (with the
Poisson ratio $\nu=0.5$). The Young's modulus of the suction cup (soft PVC) is of order (depending on the relaxation time)
$2-4 \ {\rm MPa}$, and since the contact pressure is of order $\approx 0.1 \ {\rm MPa}$, 
we are interested in $p/E$ below 0.1 and the separation at the critical constriction
in most cases is of order a few ${\rm \mu m}$. 

At the start of a pull-off experiment
the suction cup was squeezed against the PMMA surface with maximum possible hand-force (about $100 \ {\rm N}$). 
The squares in Fig. \ref{fitsuctioncup.eps}
shows the dependency of the failure time (time to pull-off) in water on the applied (pulling) force.
It is remarkable that, on the average, the failure time increases with increasing pull-off force for $F_1$ between $\approx 30 \ {\rm N}$
and $\approx 60 \ {\rm N}$ (see also dashed line in Fig. \ref{1pulloffF.2LOGlifetime.eps}).

The solid lines in Fig. \ref{fitsuctioncup.eps} are the theory predictions assuming that in the initial state, before applying the pull-off
force, there is only trapped water (but no air) (red line), or trapped air (water with an air bubble) (green line)
inside the suction cup. When only water occur, if the applied pull-force is big enough, a negative pressure (mechanical tension)
develop in the water which will pull the surfaces further together in the
contact strip of width $l(t)$ (see last term in (4)). This reduces the water leakage and increases the failure time.
When air occur in the suction cup the pressure is always positive, but below the atmospheric pressure.

In the present experiments we have used distilled but not degassed water. 
We observed that even if no air bubbles can be detected inside the suction
cup when immersed in the water, after squeezing it in contact 
with the counter surface and removing the applied squeezing force, we always observed
a cavity region (gas bubble) at the top of the suction cup. That is, 
for water with dissolved air cavitation always occurred inside the suction cup due to the reduced
pressure resulting from the spring force. We observed this even for a smooth 
glass substrate where no trapped micro or nano bubbles of gas is expected before application of the suction cup.
(For the sandblasted PMMA, gas (air) may be trapped in roughness cavities.) Consequently, some air is always trapped inside the
suction cup before application of the pull-off force $F_1$. 
Thus, the experimental results should be compared to the green line which assumes trapped air.
The deviation between theory and experiments, observed mainly for applied forces above $30 \ {\rm N}$, may be due to capillary forces. 
 
Fig. \ref{1pulloffF.2LOGlifetime.eps} 
shows the  dependency of the logarithm of the failure time (time to pull-off) on the applied (pulling) force 
in water (green squares, from Fig. \ref{fitsuctioncup.eps}) and in the air (blue squares). 
The green and blue solid lines are the theory predictions assuming that in the initial state, before applying the pull-off
force, there is some air inside the suction cup. 
The black dashed line is a fit-line to the experimental data in water.

In the air the failure time
decreased from $\approx 10 \ {\rm s}$ for the pull-off $F_1=1.4 \ {\rm N}$ to $\approx 5 \ {\rm s}$ for $F_1=10 \ {\rm N}$.
In water the failure time was typically $\approx 100$ times longer (see Fig. \ref{1pulloffF.2LOGlifetime.eps}). This is also predicted
by the theory (solid lines in \ref{1pulloffF.2LOGlifetime.eps}) and is mainly
due to the change in the viscosity which amount to a factor 
of $\approx 56$ ($\eta \approx 1.0 \times 10^{-3} \ {\rm Pa s}$ for water
and $\approx 1.8 \times10^{-5} \ {\rm Pa s}$ for air). 
The factor $Q$, which is due to the finite compressibility of air and to
ballistic air flow, is close to unity in the present case. Since the rubber-substrate contact time is longer in water, 
and since the relaxation modulus decreases with increasing time (see Ref. \cite{tobe}), in water the surfaces approach each other
more closely in the contact strip $l(t)$ then in air, which also tend to increase the failure time in water as compared to in air.

To get a deeper understanding of the failure process, in Fig. \ref{1F1.2pb.1.eps}
we show the pressure $p_{\rm b}$ in the water inside the suction cup as a function of the applied
pulling force. The vertical lines give the water pressure as the time changes from start
of pull-off ($t=0$) to detachment ($t=t_{\rm pull-off}$). The red lines assumes only water inside the
suction cup, while the green lines assumes a small air bubble trapped in the initial state. 
In the latter case the water (and air) pressure inside the suction cup is always above the vacuum pressure $p=0$,
while in the first case, when the pull-off force is bigger than
$\approx 15 \ {\rm N}$ the water pressure becomes negative for some initial time period.

Fig. \ref{1F1.2uc.1.eps}
shows the separation at the critical junction, $u_{\rm c}$, as a function of the applied
pulling force. The vertical lines give the critical separation as the time changes from start
of pull-off ($t=0$) to detachment ($t=t_{\rm pull-off}$). Again the red lines assumes only water inside the
suction cup, while the green lines assumes a small air bubble trapped in the initial state. 
The critical separation for short contact times is smaller in the former case owing to the lower
fluid pressure in the suction cup, which pull the surfaces in the nominal contact strip $l(t)$
closer together (see last term in (4)).

Fig. \ref{1pullforce.2loglifetime.1.eps}
shows a schematic picture illustrating the influence of cavitation on the failure time.
For small applied pulling force the pressure in the water in the suction cup is not low enough to
induce cavitation, but for large enough pulling-force cavitation occur. In this case,
since cavitation is a stochastic process involving thermally activated nucleation of an air bubble, 
when the experiment is repeated many times large fluctuations in the failure time may occur (dotted region).

\vskip 0.3cm
{\bf 7 Discussion}

Many animals have developed suction cups to adhere to different surfaces in water, e.g., the octopus\cite{Smith,Gorb1,Gorb2,Benne} or
northern clingfish\cite{Gorb3}. Studies have shown that these animals can adhere to much rougher
surfaces then man-made suction cups. This is due to the very low elastic modulus of the material
covering the suction cup surfaces. Thus, while most man-made suction cups are made from
rubber-like materials with a Young's modulus of order a few MPa, the suction cups of the octopus and the northern clingfish
are covered by very soft materials with an effective modulus of order $10 \ {\rm kPa}$. 

When a block of a very soft material is squeezed against a counter surface in water it tends to trap islands of water
which reduces the contact area and the friction\cite{Mug1,Mug2,Mug3}. For this reason 
the surfaces of the soft adhesive discs in the octopus and the northern clingfish 
have channels which allow the water to get removed faster during the approach 
of the suction cup to the counter surface. However, due to the low elastic modulus of the suction cup material,
the channels are ``flattened-out'' when the suction cup is in adhesive contact with the counter surface, and negligible
fluid leakage is expected to result from these surface structures.

There are two ways to attach a suction cup to a counter surface. Either a squeezing force is applied, or a
pump must be used to lower the fluid (gas or liquid) pressure inside the suction cup. The latter is used in some engineering applications.
However, it is not always easy for the octopus to apply a large normal force when attaching a suction cup to a counter surface,
in particular before any arm is attached, and when the animal cannot wind the arm around the counter surface as may be the 
case in some accounts with sperm whale.
Similarly, the northern clingfish cannot apply a large normal force to squeeze the adhesive disk in contact with a counter
surface. So how can they attach the suction cups? We believe it may be due to changes in the suction cup volume involving muscle contraction as discussed in Ref. \cite{tobe}.

For adhesion to very rough surfaces, the part of the suction cup in contact with the substrate 
must be made from an elastically very soft material. 
However, using a very soft material everywhere result in a very small suction cup stiffness. We have shown 
(see in Ref. \cite{tobe}) that a long lifetime require a large enough suction cup stiffness. 
Only in this case will the contact pressure $p$ be large enough to reduce the water
leak-rate to small enough values. Based on this, we have proposed a biomimetic design of an artificial  suction cup 
having a elastically stiff membrane covered with a soft layer with a potential to  stick to very rough surfaces under water, see  ref. \cite{tobe} for more details.

Recently two groups \cite{Dit,Sand}, working on manufacturing of suction cups 
inspired from Northern clingfish, were successful in attaining high pull-off forces for rough surfaces in water.
These devices utilize a relatively stiff material for the suction cup chamber, and a soft layer at the disc rim 
(with and without hierarchical structures), which increases the contact area with rough surfaces, 
and reduce the leakage of the fluid into the suction cups. 
Sandoval et. al.\cite{Sand} also varied the design of suction cups, 
which included radial slits to remove water from the contact. 
It was suggested\cite{Dit} that the use soft layer increases the friction on a rough substrate, 
and that this helps in reducing leakage. However, we believe that friction in itself is not very important,
but the elastically soft layer reduces the surface separation and the leakage across the interface.

For suction cups used in the air the lowest possible pressure inside the suction cup is $p_{\rm b} = 0$, 
corresponding to perfect vacuum. In reality it is usually much larger, $p_{\rm b} \approx 30-90 \ {\rm kPa}$. For suction cups
in water the pressure inside the suction cup could be negative where the liquid is under mechanical tension.
This state is only metastable, and negative pressures have been observed
for short times\cite{Cav}. For water in equilibrium with the normal atmosphere one expect cavitation to occur for any pressure below the atmospheric pressure, but the nucleation of cavities may take long time, and depends strongly on impurities
and imperfections. For example, crack-like surface defects in hydrophobic materials may trap small (micrometer or nanometer)
air bubbles which could expand to macroscopic size when the pressure is reduced below the atmospheric pressure.

Negative pressures have been observed inside the suction cups of octopus. Thus, in one study\cite{Smith1} it was found that
most suction cups have pressures above zero, but some suction cups showed pressures as low as $\approx -650 \ {\rm kPa}$.
We find this observation remarkable because for the suction cups we studied cavitation is always observed and the water pressure
is therefore always positive.

Trapped air bubbles could influence the water flow into the suction cup by blocking flow channels.
For not degassed water, whenever the fluid pressure falls below the atmospheric pressure cavitation can occur, 
and gas bubbles could form in the flow channels and block the fluid flow due to the Laplace pressure effect. 
For the suction cups studied above, in the initial state the Laplace pressure is likely to be smaller than the fluid pressure difference
between inside and outside the suction cup, in which case the gas bubbles would get removed, but at a later stage in the
detachment process this may no longer be true. This is similar to observations in earlier model studies of the water leakage of
rubber seals, where strongly reduced leakage rates was observed for hydrophobic surfaces when the  
water pressure difference between inside and outside the seal become small enough\cite{seal4}.

\vskip 0.1cm
{\bf 8 Summary and conclusion}

We have studied the leakage of suction cups both in air and water. The experimental 
results were analyzed using a newly developed theory of fluid leakage valid in diffusive and ballistic limits
combined with Persson contact mechanics theory.
In these experiments the suction cups (made of soft PVC) were pressed against sandblasted PMMA sheets. 
We found that the measured failure times of suction cups in air to be in good agreement with the theory, 
except for surfaces with rms-roughness below $\approx 1 \ {\rm \mu m}$, where diffusion of plasticizer 
occurred, from the PVC to the PMMA counterface resulting in blocking of critical constrictions.
For experiments in water, we found that the failure times of suction cup were $\approx$100 times longer than in air, and this could be attributed mainly to the different viscosity of air and water.

\bibliographystyle{apa}

\end{document}